\documentclass{ws-procs9x6}

\begin{document}

\title{A New Spectral Cancellation in Quantum Gravity}

\author{Giampiero Esposito}

\address{INFN, Sezione di Napoli, Complesso Universitario di Monte S. Angelo,
Via Cintia, Edificio N', 80126 Napoli, Italy\\
Universit\`a degli Studi di Napoli Federico II, Dipartimento di Scienze
Fisiche, Complesso Universitario di Monte S. Angelo, Via Cintia,
Edificio N', 80126 Napoli,
Italy\\E-mail: giampiero.esposito@na.infn.it}

\author{Guglielmo Fucci}

\address{Department of Physics, New Mexico Institute of Mining and
Technology, Leroy Place 801, Socorro, NM 87801, 
USA\\E-mail: gfucci@nmt.edu}

\author{Alexander Kamenshchik}

\address{Dipartimento di Fisica and INFN, Via Irnerio 46, 40126 Bologna, Italy\\
L.D. Landau Institute for Theoretical Physics, Kosygin str. 2, 119334 Moscow,
Russia\\E-mail: alexander.kamenshchik@bo.infn.it}

\author{Klaus Kirsten}

\address{Department of Mathematics, Baylor University, Waco TX 76798, 
USA\\E-mail: Klaus\_Kirsten@baylor.edu}

\maketitle

\abstracts{A general method exists for studying Abelian and
non-Abelian gauge theories, as well as Euclidean quantum gravity,
at one-loop level on manifolds with boundary. In the latter case,
boundary conditions on metric perturbations $h$ can be chosen to
be completely invariant under infinitesimal diffeomorphisms, to
preserve the invariance group of the theory and BRST symmetry. In 
the de Donder gauge, however, the resulting boundary-value problem for
the Laplace type operator acting on $h$ is 
known to be self-adjoint but not strongly elliptic.
The present paper shows that, on the Euclidean four-ball,
only the scalar part of perturbative modes
for quantum gravity is affected by the lack of strong ellipticity.
Interestingly, three sectors of the
scalar-perturbation problem remain elliptic, while lack of strong
ellipticity is ``confined'' to the remaining fourth sector.
The integral representation of the resulting $\zeta$-function 
asymptotics on the Euclidean four-ball is also obtained; this 
remains regular at the origin by virtue of a peculiar 
spectral identity obtained by the authors. There is therefore encouraging 
evidence in favour of the $\zeta(0)$ value 
with fully diff-invariant boundary conditions remaining well defined, 
at least on the four-ball, although severe technical obstructions
remain in general.
\newline
2000 {\it MSC}. Primary 58J35, 83C45; Secondary 81S40, 81T20.}

\section{Introduction}

This paper is motivated by the authors' struggle over many years with an
important problem in quantum field theory and spectral geometry, i.e. the
functional determinant in Euclidean quantum gravity
on manifolds with non-empty boundary. The related open 
issues are not yet settled, but there is a sufficient amount of new  
calculations to justify further efforts, as we are going to see shortly.

The subject of boundary effects in quantum field theory 
(Deutsch and Candelas [\refcite{1}]) has always 
received a careful consideration in the literature by virtue of very
important physical and mathematical motivations, that can be
summarized as follows.

(i) Boundary data play a crucial role in the functional-integral approach
(DeWitt [\refcite{2}]), in the quantum theory of the early universe
(Hartle and Hawking, Hawking [\refcite{3}]) in supergravity 
(Hawking [\refcite{4}]) and even in
string theory (Abouelsaood et al. [\refcite{5}]).

(ii) The way in which quantum fields react to the presence of boundaries
is responsible for remarkable physical effects, e.g. the attractive Casimir
force among perfectly conducting parallel plates (Bordag et al., 
Milton, Nesterenko et al. [\refcite{6}]), which can be viewed as
arising from differences of zero-point energies of the quantized 
electromagnetic field.

(iii) The spectral geometry of a Riemannian manifold 
(Gilkey [\refcite{7}]) with boundary is a
fascinating problem where many new results have been derived over the last
few years (Kirsten [\refcite{8}], Vassilevich [\refcite{9}]).

(iv) Boundary terms (Moss [\refcite{10}]) 
in heat-kernel expansions have become a major subject
of investigation in quantum gravity (Avramidi [\refcite{11}]), 
since they shed new light on one-loop conformal anomalies (Esposito et
al., Moss and Poletti [\refcite{12}], Tsoupros [\refcite{13}])
and one-loop divergences (Esposito [\refcite{14}], Esposito et al.
[\refcite{15}]).

In our paper we are interested in boundary 
conditions for metric perturbations that are completely invariant 
under infinitesimal diffeomorphisms, since they are part of the
general scheme according to which the boundary conditions are preserved
under the action of the symmetry group of the theory (Barvinsky 
[\refcite{16}], 
Moss and Silva [\refcite{17}], Avramidi and Esposito [\refcite{18}]). 
In field-theoretical language, this means setting to zero at the
boundary that part $\pi A$ of the gauge field $A$ that lives on
the boundary ${\mathcal B}$ ($\pi$ being a projection operator):
\begin{equation}
\Bigr[\pi A \Bigr]_{\mathcal B}=0,
\label{(1)}
\end{equation}
as well as the gauge-fixing functional,
\begin{equation}
\Bigr[\Phi(A)\Bigr]_{\mathcal B}=0,
\label{(2)}
\end{equation}
and the whole ghost field
\begin{equation}
[\varphi]_{\mathcal B}=0.
\label{(3)}
\end{equation}

For Euclidean quantum gravity, Eq. (1) reads as
\begin{equation}
[h_{ij}]_{\mathcal B}=0,
\label{(4)}
\end{equation}
where $h_{ij}$ are perturbations of the induced three-metric. 
To arrive at the gravitational 
counterpart of Eqs. (2) and (3), note first that, 
under infinitesimal diffeomorphisms, metric
perturbations $h_{\mu \nu}$ transform according to
\begin{equation}
{\widehat h}_{\mu \nu} \equiv h_{\mu \nu}+
\nabla_{(\mu} \; \varphi_{\nu)},
\label{(5)}
\end{equation}
where $\nabla$ is the Levi--Civita connection on the background
four-geometry with metric $g$, and $\varphi_{\nu}dx^{\nu}$ is
the ghost one-form (strictly, our presentation is simplified:
there are two independent ghost fields obeying Fermi statistics,
and we will eventually multiply by $-2$ the effect of $\varphi_{\nu}$
to take this into account). In geometric language, the infinitesimal
variation $\delta h_{\mu \nu} \equiv {\widehat h}_{\mu \nu}
-h_{\mu \nu}$ is given by the Lie derivative along $\varphi$ of the
four-metric $g$. For manifolds with boundary, 
Eq. (5) implies that (Esposito et al. [\refcite{19}], Avramidi et al.
[\refcite{20}])
\begin{equation}
{\widehat h}_{ij}=h_{ij}+\varphi_{(i \mid j)}
+K_{ij}\varphi_{0},
\label{(6)}
\end{equation}
where the stroke denotes three-dimensional covariant differentiation
tangentially with respect to the intrinsic Levi--Civita connection 
of the boundary, while $K_{ij}$ is the extrinsic-curvature tensor of
the boundary. Of course, $\varphi_{0}$ and $\varphi_{i}$ are the 
normal and tangential components of the ghost, respectively. By virtue
of Eq. (6), the boundary conditions (4) are ``gauge invariant'', i.e.
\begin{equation}
\Bigr[{\widehat h}_{ij}\Bigr]_{\mathcal B}=0,
\label{(7)}
\end{equation}
if and only if the whole ghost field obeys homogeneous Dirichlet
conditions, so that
\begin{equation}
[\varphi_{0}]_{\mathcal B}=0,
\label{(8)}
\end{equation}
\begin{equation}
[\varphi_{i}]_{\mathcal B}=0.
\label{(9)}
\end{equation}
The conditions (8) and (9) are necessary and sufficient since 
$\varphi_{0}$ and $\varphi_{i}$ are independent, and three-dimensional
covariant differentiation commutes with the operation of restriction
to the boundary. We are indeed assuming that the boundary 
$\mathcal B$ is smooth and not totally geodesic, i.e.
$K_{ij} \not = 0$. However, for totally geodesic boundaries, having
$K_{ij}=0$, the condition (8) is no longer necessary. 

On imposing boundary conditions on the remaining set of metric
perturbations, the key point is to make sure that {\it the invariance
of such boundary conditions under the infinitesimal transformations}
(5) {\it is again guaranteed by} (8) {\it and} (9), since otherwise
one would obtain incompatible sets of boundary conditions on the
ghost field. Indeed, on using the DeWitt--Faddeev--Popov formalism
for the $\langle {\rm out}| {\rm in} \rangle$ amplitudes of quantum
gravity, it is necessary to use a gauge-averaging term in the
Euclidean action, of the form$^{2}$
\begin{equation}
I_{g.a.}={1\over 16 \pi G}\int_{{\mathcal M}}
{\Phi_{\nu}\Phi^{\nu}\over 2\alpha}\sqrt{{\rm det} \; g} \; d^{4}x,
\label{(10)}
\end{equation}
where $\Phi_{\nu}$ is any functional which leads to self-adjoint 
(elliptic) operators on metric and ghost perturbations. One then
finds that
\begin{equation}
\delta \Phi_{\mu}(h) \equiv \Phi_{\mu}(h)-
\Phi_{\mu}({\widehat h})={\mathcal F}_{\mu}^{\; \nu} \;
\varphi_{\nu},
\label{(11)}
\end{equation}
where ${\mathcal F}_{\mu}^{\; \nu}$ is an elliptic operator that acts
linearly on the ghost field. Thus, if one imposes the boundary
conditions
\begin{equation}
\Bigr[\Phi_{\mu}(h)\Bigr]_{\mathcal B}=0,
\label{(12)}
\end{equation}
and if one assumes that the ghost field can be expanded in a
complete orthonormal set of eigenfunctions $u_{\nu}^{\; (\lambda)}$
of ${\mathcal F}_{\mu}^{\; \nu}$ which vanish at the boundary, i.e.
\begin{equation}
{\mathcal F}_{\mu}^{\; \nu} u_{\nu}^{\; (\lambda)}=\lambda
u_{\mu}^{\; (\lambda)},
\label{(13)}
\end{equation}
\begin{equation}
\varphi_{\nu}=\sum_{\lambda}C_{\lambda}u_{\nu}^{\; (\lambda)},
\label{(14)}
\end{equation}
\begin{equation}
\Bigr[u_{\mu}^{\; (\lambda)}\Bigr]_{\mathcal B}=0,
\label{(15)}
\end{equation}
the boundary conditions (12) are automatically gauge-invariant under
the Dirichlet conditions (8) and (9) on the ghost.

Having obtained the general recipe expressed by Eqs. (4) and (12), we
can recall what they imply on the Euclidean four-ball. This background
is relevant for one-loop quantum cosmology in the limit of small
three-geometry on the one hand (Schleich [\refcite{21}]), 
and for spectral geometry and spectral
asymptotics on the other hand [\refcite{8,9}]. As shown in Ref. 19, 
if one chooses the de Donder gauge-fixing functional
\begin{equation}
\Phi_{\mu}(h)=\nabla^{\nu}\Bigr(h_{\mu \nu}-{1\over 2}
g_{\mu \nu}g^{\rho \sigma}h_{\rho \sigma}\Bigr),
\label{(16)}
\end{equation}
which has the virtue of leading to an operator of Laplace type on
$h_{\mu \nu}$ in the one-loop functional integral, Eq. (12) yields 
the mixed boundary conditions
\begin{equation}
\left[{\partial h_{00}\over \partial \tau}+{6\over \tau}h_{00}
-{\partial \over \partial \tau}(g^{ij}h_{ij})
+{2\over \tau^{2}}h_{0i}^{\; \mid i}\right]_{\mathcal B}=0,
\label{(17)}
\end{equation}
\begin{equation}
\left[{\partial h_{0i}\over \partial \tau}+{3\over \tau}h_{0i}
-{1\over 2}{\partial h_{00}\over \partial x^{i}}
\right]_{\mathcal B}=0.
\label{(18)}
\end{equation}
In Refs. 15, 19, 
the boundary conditions (4), (17) and (18) were 
used to evaluate the full one-loop divergence of quantized general
relativity on the Euclidean four-ball, including all $h_{\mu \nu}$
and all ghost modes. However, the meaning of such a calculation
became unclear after the discovery in Ref. 18 that the 
boundary-value problem for the Laplacian $P$ acting on metric
perturbations is not strongly elliptic by virtue of tangential
derivatives in the boundary conditions (17) and (18). Moreover,
the work by Dowker and Kirsten [\refcite{22}] had proved even earlier,
in a simpler case, that the boundary-value problem with tangential
derivatives is, in general, not strongly elliptic. Strong
ellipticity [\refcite{8,18}] 
is a technical requirement ensuring that a unique
smooth solution of the boundary-value problem exists which vanishes
at infinite geodesic distance from the boundary. 
If it is fulfilled,
this ensures that the $L^{2}$ trace of the heat semigroup $e^{-tP}$ 
exists, with the associated global heat-kernel asymptotics that 
yields one-loop divergence and one-loop effective action. However,
when strong ellipticity does not hold, the $L^{2}$ trace of
$e^{-tP}$ acquires a singular part [\refcite{18}] and hence 
$\zeta$-function calculations may become ill-defined. 

All of this has motivated our analysis, which therefore derives in Sec.
2 the eigenvalue conditions for scalar modes. Section 3
obtains the first pair of resulting scalar-mode 
$\zeta$-functions and Sec. 4 studies the remaining
elliptic and non-elliptic parts of spectral asymptotics. Results and
open problems are described in Sec. 5.

\section{Eigenvalue conditions for scalar modes on the four-ball}

On the Euclidean four-ball, which can be viewed as the portion of
flat Euclidean four-space bounded by a three-sphere of radius $q$,
metric perturbations $h_{\mu \nu}$ can be
expanded in terms of hyperspherical harmonics as (Lifshitz and Khalatnikov 
[\refcite{23}], Esposito et al. [\refcite{24}])
\begin{equation}
h_{00}(x,\tau)=\sum_{n=1}^{\infty}a_{n}(\tau)Q^{(n)}(x),
\label{(19)}
\end{equation}
\begin{equation}
h_{0i}(x,\tau)=\sum_{n=2}^{\infty}\left[b_{n}(\tau)
{Q_{\mid i}^{(n)}(x)\over (n^{2}-1)}+c_{n}(\tau)S_{i}^{(n)}(x)\right],
\label{(20)}
\end{equation}
\begin{eqnarray}
h_{ij}(x,\tau)&=& \sum_{n=3}^{\infty}d_{n}(\tau)\left[
{Q_{\mid ij}^{(n)}(x)\over (n^{2}-1)}+{c_{ij}\over 3}
Q^{(n)}(x)\right]+\sum_{n=1}^{\infty}{e_{n}(\tau)\over 3}c_{ij}
Q^{(n)}(x) \nonumber \\
&+& \sum_{n=3}^{\infty}\left[f_{n}(\tau)\Bigr(S_{i \mid j}^{(n)}(x)
+S_{j \mid i}^{(n)}(x) \Bigr)+k_{n}(\tau)G_{ij}^{(n)}(x)\right],
\label{(21)}
\end{eqnarray}
where $\tau \in [0,q]$ and 
$Q^{(n)}(x), S_{i}^{(n)}(x)$ and $G_{ij}^{(n)}(x)$ are scalar,
transverse vector and transverse-traceless tensor hyperspherical
harmonics, respectively, on a unit three-sphere with metric $c_{ij}$. 
By insertion of the expansions (19)-(21) into the eigenvalue equation
for the Laplacian acting on $h_{\mu \nu}$, and by setting 
$\sqrt{E} \rightarrow iM$, which corresponds to a rotation of
contour in the $\zeta$-function analysis (Barvinsky et al.  
[\refcite{25}]) one finds the modes as
linear combinations of modified Bessel functions of first kind. 
Modified Bessel functions of the second kind are not included to
ensure regularity at the origin $\tau=0$. For details, we refer the
reader to the work by Esposito et al. [\refcite{26}].

The boundary conditions (4), (17), (18), (8), (9), jointly with
the mode-expansions on the four-ball, can be used to obtain
homogeneous linear systems that yield, implicitly, the eigenvalues
of our problem. The conditions for finding non-trivial solutions 
of such linear systems are given by the vanishing of the associated
determinants; these yield the eigenvalue conditions
$\delta(E)=0$, i.e. the equations obeyed by the eigenvalues by
virtue of the boundary conditions. For the purpose of a
rigorous analysis, we need the full expression of such
eigenvalue conditions for each set of coupled modes. 
Upon setting $\sqrt{E} \rightarrow iM$, we denote by $D(Mq)$ the
counterpart of $\delta(E)$, bearing in mind that, strictly,
only $\delta(E)$ yields implicitly the eigenvalues, while
$D(Mq)$ is more convenient for $\zeta$-function calculations
[\refcite{25}].

In particular, we here focus on scalar modes (for the whole set of
modes, see again the work in Ref. 26).
For all $n \geq 3$, coupled scalar modes $a_{n},b_{n},d_{n},e_{n}$ 
are ruled by a determinant reading as
\begin{equation}
D_{n}(Mq)={\rm det} \; \rho_{ij}(Mq),
\label{(22)}
\end{equation}
with degeneracy $n^{2}$, where $\rho_{ij}$ is a $4 \times 4$ 
matrix with entries (hereafter, $I_{n}$ are modified Bessel functions
of first kind)
\begin{equation}
\rho_{11}=I_{n}(Mq)-Mq I_{n}'(Mq), \;
\rho_{12}=Mq I_{n}'(Mq),
\label{(23)}
\end{equation}
\begin{eqnarray}
\; & \; & \rho_{13}=(2-n)I_{n-2}(Mq)+MqI_{n-2}'(Mq), \nonumber \\
& \; & \rho_{14}=(2+n)I_{n+2}(Mq)+MqI_{n+2}'(Mq),
\label{(24)}
\end{eqnarray}
\begin{equation}
\rho_{21}=-(n^{2}-1)I_{n}(Mq), \;
\rho_{22}=2MqI_{n}'(Mq)+6I_{n}(Mq),
\label{(25)}
\end{equation}
\begin{equation}
\rho_{23}=2(n+1)MqI_{n-2}'(Mq)-(n^{2}-6n-7)I_{n-2}(Mq),
\label{(26)}
\end{equation}
\begin{equation}
\rho_{24}=-2(n-1)MqI_{n+2}'(Mq)-(n^{2}+6n-7)I_{n+2}(Mq),
\label{(27)}
\end{equation}
\begin{equation}
\rho_{31}=0, \;
\rho_{32}=-I_{n}(Mq),
\label{(28)}
\end{equation}
\begin{equation}
\rho_{33}={(n+1)\over (n-2)}I_{n-2}(Mq), \;
\rho_{34}={(n-1)\over (n+2)}I_{n+2}(Mq),
\label{(29)}
\end{equation}
\begin{equation}
\rho_{41}=3I_{n}(Mq), \;
\rho_{42}=-2I_{n}(Mq), \;
\rho_{43}=-I_{n-2}(Mq), \;
\rho_{44}=-I_{n+2}(Mq).
\label{(30)}
\end{equation}
The hardest part of our analysis is the investigation of 
the equation obtained by setting to zero the determinant (22).
For this purpose, we first exploit the recurrence relations 
among $I_{n},I_{n+1}$ and $I_{n}'$ to find (from now on,
$w \equiv Mq$)
\begin{eqnarray}
\; & \; & \rho_{11}=I_{n}(w)-wI_{n}'(w), \; 
\rho_{12}=wI_{n}'(w), \; 
\rho_{13}=wI_{n}'(w)+nI_{n}(w), \nonumber \\
& \; & \rho_{14}=wI_{n}'(w)-nI_{n}(w),
\label{(31)}
\end{eqnarray}
\begin{equation}
\rho_{21}=-(n^{2}-1)I_{n}(w), \;
\rho_{22}=2(wI_{n}'(w)+3I_{n}(w)),
\label{(32)}
\end{equation}
\begin{eqnarray}
\rho_{23}&=&(n+1)\biggr \{ \left[3(n+1)+{2n(n-1)(n+3)\over w^{2}}\right]
I_{n}(w) \nonumber \\
&+& 2 \left[w+{(n-1)(n+3)\over w}\right]I_{n}'(w) \biggr \},
\label{(33)}
\end{eqnarray}
\begin{eqnarray}
\rho_{24}&=&
(n-1)\biggr \{ \left[3(n-1)+{2n(n+1)(n-3)\over w^{2}}\right]
I_{n}(w) \nonumber \\
&-& 2 \left[w+{(n+1)(n-3)\over w}\right]I_{n}'(w) \biggr \},
\label{(34)}
\end{eqnarray}
\begin{equation}
\rho_{31}=0, \; \rho_{32}=-I_{n}(w),
\label{(35)}
\end{equation}
\begin{equation}
\rho_{33}={(n+1)\over (n-2)}\left[\left(1+{2n(n-1)\over w^{2}}
\right)I_{n}(w)+{2(n-1)\over w}I_{n}'(w)\right],
\label{(36)}
\end{equation}
\begin{equation}
\rho_{34}={(n-1)\over (n+2)}\left[\left(1+{2n(n+1)\over w^{2}}
\right)I_{n}(w)-{2(n+1)\over w}I_{n}'(w)\right],
\label{(37)}
\end{equation}
\begin{equation}
\rho_{41}=3I_{n}(w), \; \rho_{42}=-2I_{n}(w),
\label{(38)}
\end{equation}
\begin{equation}
\rho_{43}=-\left(1+{2n(n-1)\over w^{2}}\right)I_{n}(w)
-{2(n-1)\over w}I_{n}'(w),
\label{(39)}
\end{equation}
\begin{equation}
\rho_{44}=-\left(1+{2n(n+1)\over w^{2}}\right)I_{n}(w)
+{2(n+1)\over w}I_{n}'(w).
\label{(40)}
\end{equation}
The resulting determinant, despite its cumbersome expression, 
can be studied by introducing the variable
\begin{equation}
y \equiv {I_{n}'(w)\over I_{n}(w)},
\label{(41)}
\end{equation}
which leads to 
\begin{equation}
D_{n}(w)={48n(1-n^{2})\over (n^{2}-4)}
I_{n}^{4}(w)(y-y_{1})(y-y_{2})(y-y_{3})(y-y_{4}),
\label{(42)}
\end{equation}
where
\begin{equation}
y_{1} \equiv -{n\over w}, \;
y_{2} \equiv {n\over w}, \;
y_{3} \equiv -{n \over w}-{w \over 2}, \;
y_{4} \equiv {n \over w}-{w \over 2},
\label{(43)}
\end{equation}
and hence
\begin{eqnarray}
\; & \; & {(n^{2}-4)\over 48n(1-n^{2})}D_{n}(w)
=\left(I_{n}'(w)+{n \over w}I_{n}(w)\right)
\left(I_{n}'(w)-{n \over w}I_{n}(w)\right) \nonumber \\
& \cdot & \Bigr(I_{n}'(w)+\Bigr({w \over 2}+{n \over w}
\Bigr)I_{n}(w)\Bigr)
\Bigr(I_{n}'(w)+\Bigr({w \over 2}-{n \over w}
\Bigr)I_{n}(w)\Bigr).
\label{(44)}
\end{eqnarray}

\section{First pair of scalar-mode $\zeta$-functions}

In our problem, the differential operator 
under investigation is the Laplacian on the Euclidean 
four-ball acting on metric perturbations. The boundary conditions
for vector, tensor and ghost modes 
correspond to a familiar mixture of Dirichlet and Robin boundary
conditions for which an integral representation of the
$\zeta$-function and heat-kernel coefficients are immediately obtained.
New features arise instead from Eq. (44), that gives rise
to four different $\zeta$-functions.
On studying the first line of Eq. (44), we exploit the
Cauchy integral formula to express the power $-s$ of the eigenvalues
and hence turn the $\zeta$-function
$$
\zeta_{A}^{\pm}(s) \equiv \sum_{n=3}^{\infty}n^{2}
\lambda_{A^{\pm}}^{-s}
$$
into an integral, i.e. we use
$$
\sum_{l=1}^{\infty}x_{l}^{-s}=\int_{\gamma}dx \; x^{-s}
{d\over dx}\log H_{n}(x),
$$
where $\gamma$ encloses the zeros $x_{1},x_{2},...,x_{\infty}$ of the
function $H_{n}$, which here equals $J_{n}'(x) \pm {n\over x}J_{n}(x)$.  
Such a combination of $J_{n}$ and $J_{n}'$ is proportional
to the power of degree $(\beta_{\pm}-1)$ of the independent variable
multiplied by an infinite product, with $\beta_{+}(n) \equiv n,
\beta_{-}(n) \equiv n+2$. Only the infinite product encodes information on 
the countable infinity of non-vanishing zeros, and hence one should
divide $xJ_{n}'(x)\pm n J_{n}(x)$ by $x^{\beta_{\pm}}$. 
Last, rotation of contour to the imaginary axis 
(Dowker and Kirsten [\refcite{22}],
Bordag et al. [\refcite{27}]), which brings in modified Bessel  
functions $I_{n}$, jointly with setting $w=zn$, leads
to the following integral formula: 
\begin{equation}
\zeta_{A}^{\pm}(s) \equiv {(\sin \pi s)\over \pi}\sum_{n=3}^{\infty}
n^{-(2s-2)}  
\int_{0}^{\infty}dz \; z^{-2s}{\partial \over \partial z}
{\rm log}\left[{\Bigr(znI_{n}'(zn) 
\pm n I_{n}(zn)\Bigr)\over z^{\beta_{\pm}(n)}}\right].
\label{(45)}
\end{equation}
 
The uniform asymptotic expansion of modified Bessel functions
and their first derivatives (see Appendix) can be used to find
(hereafter $\tau=\tau(z) \equiv (1+z^{2})^{-{1\over 2}}$)
\begin{equation}
znI_{n}'(zn) \pm n I_{n}(zn) \sim
{n \over \sqrt{2\pi n}}{e^{n \eta}\over \sqrt{\tau}}
(1 \pm \tau)\left(1+\sum_{k=1}^{\infty}{p_{k,\pm}(\tau)\over n^{k}}
\right),
\label{(46)}
\end{equation}
where (see Eqs. (139) and (141) in the Appendix for the functions
$u_{k}$ and $v_{k}$)
\begin{equation}
p_{k,\pm}(\tau) \equiv (1\pm \tau)^{-1}\Bigr(v_{k}(\tau) \pm 
\tau u_{k}(\tau)\Bigr),
\label{(47)}
\end{equation}
for all $k \geq 1$, and
\begin{equation}
{\rm log} \left(1+\sum_{k=1}^{\infty}
{p_{k,\pm}(\tau)\over n^{k}}\right) \sim
\sum_{k=1}^{\infty}{T_{k,\pm}(\tau)\over n^{k}}.
\label{(48)}
\end{equation}
Thus, the $\zeta$-functions (45) obtain, from the first pair of 
round brackets in Eq. (46), the contributions (cf. Ref. 22)
\begin{equation}
A_{+}(s) \equiv 
\sum_{n=3}^{\infty} n^{-(2s-2)} 
{(\sin \pi s)\over \pi}
\int_{0}^{\infty}dz \; z^{-2s}{\partial \over \partial z}
\log \Bigr(1 + (1+z^{2})^{-{1\over 2}}\Bigr),
\label{(49)}
\end{equation}
\begin{equation}
A_{-}(s) \equiv 
\sum_{n=3}^{\infty} n^{-(2s-2)} 
{(\sin \pi s)\over \pi}
\int_{0}^{\infty}dz \; z^{-2s}{\partial \over \partial z}
\log \left({1 - (1+z^{2})^{-{1\over 2}} \over z^{2}}\right),
\label{(50)}
\end{equation}
where $z^{2}$ in the denominator of the argument of the $\log$
arises, in Eq. (50), from the extra $z^{-2}$ in the prefactor
$z^{-\beta_{-}(n)}$ in the definition (45). Moreover, the
second pair of round brackets in Eq. (46) contributes
$\sum_{j=1}^{\infty}A_{j,\pm}(s)$, having defined
\begin{equation}
A_{j,\pm}(s) \equiv 
\sum_{n=3}^{\infty} n^{-(2s+j-2)}
{(\sin \pi s)\over \pi}
\int_{0}^{\infty}dz \; z^{-2s}
{\partial \over \partial z} T_{j,\pm}(\tau(z)),
\label{(51)}
\end{equation}
where, from the formulae
\begin{equation}
T_{1,\pm}=p_{1,\pm},
\label{(52)}
\end{equation}
\begin{equation}
T_{2,\pm}=p_{2,\pm}-{1\over 2}p_{1,\pm}^{2},
\label{(53)}
\end{equation}
\begin{equation}
T_{3,\pm}=p_{3,\pm}-p_{1,\pm}p_{2,\pm}+{1\over 3}p_{1,\pm}^{3},
\label{(54)}
\end{equation}
we find
\begin{equation}
T_{1,\pm}=-{3\over 8}\tau \pm{1\over 2}\tau^{2}-{5\over 24}\tau^{3},
\label{(55)}
\end{equation}
\begin{equation}
T_{2,\pm}=-{3\over 16}\tau^{2} \pm{3\over 8}\tau^{3}
+{1\over 8}\tau^{4} \mp{5\over 8}\tau^{5}
+{5\over 16}\tau^{6}, 
\label{(56)}
\end{equation}
\begin{equation}
T_{3,\pm}=-{21\over 128}\tau^{3} \pm{3\over 8}\tau^{4}
+{509\over 640}\tau^{5} \mp{25\over 12}\tau^{6}
+{21\over 128}\tau^{7} \pm{15\over 8}\tau^{8}
-{1105 \over 1152}\tau^{9},
\label{(57)}
\end{equation}
and hence, in general,
\begin{equation}
T_{j,\pm}(\tau)=\sum_{a=j}^{3j}f_{a}^{(j,\pm)}\tau^{a}.
\label{(58)}
\end{equation}

We therefore find, from the first line of Eq. (44),
contributions to the generalized $\zeta$-function, from terms in
round brackets in Eq. (46), equal to
\begin{equation}
\chi_{A}^{\pm}(s)=\omega_{0}(s)F_{0}^{\pm}(s)+\sum_{j=1}^{\infty}
\omega_{j}(s)F_{j}^{\pm}(s),
\label{(59)}
\end{equation}
where, for all $\lambda=0,j$ ($\zeta_{R}$ and $\zeta_{H}$ being the
Riemann and Hurwitz $\zeta$-functions, respectively),
\begin{equation}
\omega_{\lambda}(s) \equiv \sum_{n=3}^{\infty}n^{-(2s+\lambda-2)}
= \zeta_{H}(2s+\lambda-2;3)
=\zeta_{R}(2s+\lambda-2)-1-2^{-(2s+\lambda-2)},
\label{(60)}
\end{equation}
while, from Eqs. (49)--(51),
\begin{equation}
F_{0}^{+}(s) \equiv {(\sin \pi s)\over \pi}\int_{0}^{\infty}
dz \; z^{-2s}{\partial \over \partial z}
\log \Bigr(1+(1+z^{2})^{-{1\over 2}}\Bigr),
\label{(61)}
\end{equation}
\begin{equation}
F_{0}^{-}(s) \equiv -2{(\sin \pi s)\over \pi}\int_{0}^{\infty}dz
\; {z^{-(2s-1)} \over (1+z^{2})}-F_{0}^{+}(s)
=-1-F_{0}^{+}(s),
\label{(62)}
\end{equation}
\begin{equation}
F_{j}^{\pm}(s) \equiv {(\sin \pi s)\over \pi}
\sum_{a=j}^{3j} L^{\pm}(s,a,0)f_{a}^{(j,\pm)},
\label{(63)}
\end{equation}
having set (this general definition will prove useful later, and arises
from a more general case, where $\tau^{a}$ is divided by the
$b$-th power of $(1 \pm \tau)$ in Eq. (58))
\begin{equation}
L^{\pm}(s,a,b) \equiv \int_{0}^{1}
\tau^{2s+a}(1-\tau)^{-s}(1+\tau)^{-s}
\Bigr(\pm b (1\pm \tau)^{-b-1}-a\tau^{-1}(1\pm \tau)^{-b}
\Bigr)d\tau.
\label{(64)}
\end{equation}

Moreover, on considering
\begin{equation}
L_{0}^{+}(s) \equiv {\pi \over \sin \pi s}F_{0}^{+}(s),
\label{(65)}
\end{equation}
and changing variable from $z$ to $\tau$ therein, all $L$-type
integrals above can be obtained from
\begin{equation}
Q(\alpha,\beta,\gamma) \equiv \int_{0}^{1}\tau^{\alpha}
(1-\tau)^{\beta}(1+\tau)^{\gamma}d\tau.
\label{(66)}
\end{equation}
In particular, we will need
\begin{equation}
L_{0}^{+}(s)=-Q(2s,-s,-s-1),
\label{(67)}
\end{equation}
\begin{equation}
L^{+}(s,a,b)=bQ(2s+a,-s,-s-b-1)-aQ(2s+a-1,-s,-s-b),
\label{(68)}
\end{equation}
where, from the integral representation of the
hypergeometric function, one has (Gradshteyn and Ryzhik [\refcite{28}])
\begin{equation}
Q(\alpha,\beta,\gamma)={\Gamma(\alpha+1)\Gamma(\beta+1)\over
\Gamma(\alpha+\beta+2)}F(-\gamma,\alpha+1;\alpha+\beta+2;-1).
\label{(69)}
\end{equation}
For example, explicitly, 
\begin{equation}
L_{0}^{+}(s)=-{\Gamma(2s+1)\Gamma(1-s)\over \Gamma(s+2)}
F(s+1,2s+1;s+2;-1).
\label{(70)}
\end{equation}

Now we exploit Eqs. (45), (46) and (59) to write
\begin{eqnarray}
\zeta_{A}^{+}(s)&=&\chi_{A}^{+}(s)+{(\sin \pi s)\over \pi}
\sum_{n=3}^{\infty}n^{-(2s-2)}\int_{0}^{\infty}dz 
\biggr[{z^{-(2s-1)}\over 2(1+z^{2})} \nonumber \\
&+& nz^{-(2s+1)} \left(\sqrt{1+z^{2}}-1 \right) \biggr].
\label{(71)}
\end{eqnarray}
Hence we find
\begin{equation}
\zeta_{A}^{+}(0)=\lim_{s \to 0}\left[\omega_{0}(s)F_{0}^{+}(s)
+\sum_{j=1}^{\infty}\omega_{j}(s)F_{j}^{+}(s)
+\Bigr(\zeta_{A}^{+}(s)-\chi_{A}^{+}(s)\Bigr)\right].
\label{(72)}
\end{equation}
The first limit in Eq. (72) is immediately obtained 
by noting that
\begin{equation}
\lim_{s \to 0}
L_{0}^{+}(s)=-\log(2),
\label{(73)}
\end{equation}
and hence
\begin{equation}
\lim_{s \to 0}\omega_{0}(s)F_{0}^{+}(s)
=\lim_{s \to 0}\left[\zeta_{H}(2s-2;3){(\sin \pi s)\over \pi}
L_{0}^{+}(s)\right]=0.
\label{(74)}
\end{equation}
To evaluate the second limit in Eq. (72), we use
\begin{equation}
\lim_{s \to 0}L^{+}(s,a,0)=-1,
\label{(75)}
\end{equation}
and bear in mind  
that $\omega_{j}(s)$ is a meromorphic function with 
first-order pole, as $s \rightarrow 0$, 
only at $j=3$ by virtue of the limit
\begin{equation}
\lim_{y \to 1}\left[\zeta_{R}(y)-{1\over (y-1)}\right]=\gamma.
\label{(76)}
\end{equation}
Hence we find (see coefficients in Eq. (57)) 
\begin{eqnarray}
\lim_{s \to 0}\sum_{j=1}^{\infty}\omega_{j}(s)F_{j}^{+}(s)
&=& \lim_{s \to 0}{(\sin \pi s)\over \pi}\sum_{j=1}^{\infty}
\omega_{j}(s)\left[\sum_{a=j}^{3j}L^{+}(s,a,0)f_{a}^{(j,+)}\right]
\nonumber \\
&=& -{1\over 2}\sum_{a=3}^{9}f_{a}^{(3,+)}=-{1\over 720},
\label{(77)}
\end{eqnarray}
while, from Eqs. (71) and (69),
\begin{eqnarray}
\lim_{s \to 0}\Bigr(\zeta_{A}^{+}(s)-\chi_{A}^{+}(s)\Bigr)
&=& \lim_{s \to 0}\left({1\over 4}\zeta_{H}(2s-2;3)
+{1\over 4 \sqrt{\pi}}{\Gamma \left(s-{1\over 2}\right)
\over \Gamma(s+1)} \zeta_{H}(2s-3;3)\right) \nonumber \\
&=& -{5\over 4}+{1079\over 240} .
\label{(78)}
\end{eqnarray}

We therefore find, with the same algorithms as in Ref. 27,
\begin{equation}
\zeta_{A}^{+}(0)=-{5\over 4}+{1079\over 240}
-{1\over 2}\sum_{a=3}^{9}f_{a}^{(3,+)}
={146\over 45},
\label{(79)}
\end{equation}
\begin{equation}
\zeta_{A}^{-}(0)=-{5\over 4}+{1079\over 240}+5
-{1\over 2}\sum_{a=3}^{9}f_{a}^{(3,-)}
={757\over 90}.
\label{(80)}
\end{equation}
These results have been double-checked by using also the powerful 
analytic technique in Ref. 25.

\section{Further spectral asymptotics: elliptic and non-elliptic parts}

As a next step, the second line of Eq. (44) 
suggests considering $\zeta$-functions
having the integral representation (using again the Cauchy theorem
and rotation of contour as in Eq. (45))
\begin{eqnarray}
\; & \; & \zeta_{B}^{\pm}(s) \equiv {(\sin \pi s)\over \pi}\sum_{n=3}^{\infty}
n^{-(2s-2)} \nonumber \\
& \; & \int_{0}^{\infty}
dz \; z^{-2s}{\partial \over \partial z}{\rm log}
\left[z^{-\beta_{\pm}(n)}\left(znI_{n}'(zn)+\left({z^{2}n^{2}\over 2}
\pm n \right)I_{n}(zn)\right)\right].
\label{(81)}
\end{eqnarray}
To begin, we exploit again the uniform asymptotic 
expansion of modified
Bessel functions and their first derivatives 
to find (cf. Eq. (46))
\begin{equation}
znI_{n}'(zn)+\left({z^{2}n^{2}\over 2} \pm n \right)I_{n}(zn)
\sim {n^{2}\over 2 \sqrt{2 \pi n}}{e^{n \eta} \over \sqrt{\tau}}
\left({1\over \tau}-\tau \right) 
\left (1+ \sum_{k=1}^{\infty}{r_{k,\pm}(\tau)\over n^{k}}
\right),
\label{(82)}
\end{equation}
where we have (bearing in mind that $u_{0}=v_{0}=1$)
\begin{equation}
r_{k,\pm}(\tau) \equiv u_{k}(\tau)
+{2\tau \over (1-\tau^{2})}
\Bigr((v_{k-1}(\tau) \pm \tau 
u_{k-1}(\tau) \Bigr),
\label{(83)}
\end{equation}
for all $k \geq 1$. Hereafter we set
\begin{equation}
\Omega \equiv \sum_{k=1}^{\infty}{r_{k,\pm}(\tau(z))\over n^{k}},
\label{(84)}
\end{equation}
and rely upon the formula
\begin{equation}
\log(1+\Omega) \sim \sum_{k=1}^{\infty}(-1)^{k+1}
{\Omega^{k}\over k}
\label{(85)}
\end{equation}
to evaluate the uniform asymptotic expansion (cf. Eq. (48))
\begin{equation}
{\rm log}\left(1+\sum_{k=1}^{\infty}{r_{k,\pm}(\tau(z))\over n^{k}}
\right) \sim \sum_{k=1}^{\infty}{R_{k,\pm}(\tau(z))\over n^{k}}.
\label{(86)}
\end{equation}
The formulae yielding $R_{k,\pm}$ from $r_{k,\pm}$ are exactly as in
Eqs. (52)--(54), with $T$ replaced by $R$ and $p$ replaced by $r$
(see, however, comments below Eq. (90)).
Hence we find, bearing in mind Eq. (83),
\begin{equation}
R_{1,\pm}=(1 \mp \tau)^{-1}
\left({17\over 8}\tau \mp{1\over 8}\tau^{2}-{5\over 24}\tau^{3}
\pm{5\over 24}\tau^{4}\right),
\label{(87)}
\end{equation}
\begin{equation}
R_{2,\pm}=(1 \mp \tau)^{-2}
\left(-{47\over 16}\tau^{2} \pm{15\over 8}\tau^{3}-{21\over 16}\tau^{4}
\pm{3\over 4}\tau^{5}-{1\over 16}\tau^{6} \mp{5\over 8}\tau^{7}
+{5\over 16}\tau^{8}\right),
\label{(88)}
\end{equation}
\begin{eqnarray}
R_{3,\pm}&=& (1 \mp \tau)^{-3}\biggr(
{1721\over 384}\tau^{3} \mp{441\over 128}\tau^{4}+{597\over 320}\tau^{5}
\mp{1033\over 960}\tau^{6} 
+{239\over 80}\tau^{7} \nonumber \\
&\mp & {28\over 5}\tau^{8}
+{2431\over 576}\tau^{9} \pm{221\over 192}\tau^{10} 
- {1105\over 384}\tau^{11} \pm{1105\over 1152}\tau^{12}\biggr),
\label{(89)}
\end{eqnarray}
and therefore
\begin{equation}
R_{j,\pm}(\tau(z))=(1 \mp \tau)^{-j}\sum_{a=j}^{4j}C_{a}^{(j,\pm)}
\tau^{a},
\label{(90)}
\end{equation}
where, unlike what happens for the $T_{j,\pm}$ polynomials, 
the exponent of $(1 \mp \tau)$ never vanishes. Note that, at
$\tau=1$ (i.e. $z=0$), our $r_{k,+}(\tau)$ and $R_{k,+}(\tau)$
are singular. Such a behaviour
is not seen for any of the strongly elliptic boundary-value
problems [\refcite{8}]. This technical 
difficulty motivates our efforts below and is interpreted by us as 
a clear indication of the lack of strong ellipticity proved,
on general ground, in Ref. 18.

The $\zeta_{B}^{-}(s)$ function is more easily dealt with.
It indeed receives contributions from terms in 
round brackets in Eq. (82) equal to (cf. Eq. (50) and
bear in mind that $\beta_{-}-\beta_{+}=2$ in Eq. (81))
\begin{eqnarray}
B_{-}(s)& \equiv & \sum_{n=3}^{\infty}n^{-(2s-2)}{(\sin \pi s)\over \pi}
\int_{0}^{\infty}dz \; z^{-2s}{\partial \over \partial z}
\log \left({{1\over \tau(z)}-\tau(z) \over z^{2}}\right)
\nonumber \\
&=& \omega_{0}(s){(\sin \pi s)\over \pi}
\int_{0}^{\infty}dz \; z^{-2s}
{\partial \over \partial z} \log {1\over \sqrt{1+z^{2}}}
=-{1\over 2}\omega_{0}(s),
\label{(91)}
\end{eqnarray}
and $\sum_{j=1}^{\infty}B_{j,-}(s)$, having defined, with  
$\lambda=0,j$ (cf. Eq. (51))
\begin{equation}
\omega_{\lambda}(s) \equiv \sum_{n=3}^{\infty}n^{-(2s+\lambda-2)}
=\zeta_{H}(2s+\lambda-2;3),
\label{(92)}
\end{equation}
\begin{equation}
B_{j,-}(s) \equiv \omega_{j}(s){(\sin \pi s)\over \pi}
\int_{0}^{\infty}dz \; z^{-2s}{\partial \over \partial z}
R_{j,-}(\tau(z)).
\label{(93)}
\end{equation}
On using the same method as in Sec. 3, 
the formulae (81)--(93) lead to 
\begin{equation}
\zeta_{B}^{-}(0)=-{5\over 4}+{1079\over 240}+{5\over 2}
-{1\over 16}\sum_{a=3}^{12}C_{a}^{(3,-)} 
={206\over 45},
\label{(94)}
\end{equation}
a result which agrees with a derivation of $\zeta_{B}^{-}(0)$ 
relying upon the method of Ref. 25.

Although we have stressed after Eq. (90) 
the problems with the $\zeta_{B}^{+}(s)$
part, for the moment let us proceed formally in the same way
as above. Thus we define, in analogy to Eq. (91),
\begin{equation}
B_{+}(s)\equiv \omega_{0}(s){(\sin \pi s)\over \pi}
\int_{0}^{\infty}dz \; z^{-2s}
{\partial \over \partial z} \log \left({1\over \tau(z)}
-\tau(z)\right),
\label{(95)}
\end{equation}
and, in analogy to Eq. (93),
\begin{equation}
B_{j,+}(s) \equiv \omega_{j}(s){(\sin \pi s)\over \pi}
\int_{0}^{\infty}dz \; z^{-2s}{\partial \over \partial z}
R_{j,+}(\tau(z)).
\label{(96)}
\end{equation}
In order to make the presentation as transparent as possible, we
write out the derivatives of $R_{j,+}$. On changing integration
variable from $z$ to $\tau$ we define
\begin{equation}
C_{j}(\tau) \equiv {\partial \over \partial \tau}R_{j,+}(\tau),
\label{(97)}
\end{equation}
and we find the following results:
\begin{equation}
C_{1}(\tau)=
(1-\tau)^{-2}\left({17\over 8}-{1\over 4}\tau-{1\over 2}\tau^{2}
+{5\over 4}\tau^{3}-{5\over 8}\tau^{4}\right),
\label{(98)}
\end{equation}
\begin{eqnarray}
C_{2}(\tau)&=&
(1-\tau)^{-3}\biggr(-{47\over 8}\tau+{45\over 8}\tau^{2}
-{57\over 8}\tau^{3}+{51\over 8}\tau^{4}-{21\over 8}\tau^{5}
-{33\over 8}\tau^{6}+{45\over 8}\tau^{7} \nonumber \\
&-& {15\over 8}\tau^{8} \biggr),
\label{(99)}
\end{eqnarray}
\begin{eqnarray}
C_{3}(\tau)&=&
(1-\tau)^{-4}\biggr({1721\over 128}\tau^{2}-{441\over 32}\tau^{3}
+{1635\over 128}\tau^{4}-{163\over 16}\tau^{5}
+{1545\over 64}\tau^{6}-{227\over 4}\tau^{7} \nonumber \\
&+& {4223\over 64}\tau^{8}-{221\over 16}\tau^{9}
-{5083\over 128}\tau^{10}+{1105\over 32}\tau^{11}
-{1105\over 128}\tau^{12}\biggr),
\label{(100)}
\end{eqnarray}
so that the general expression of $C_{j}(\tau)$ reads as
\begin{equation}
C_{j}(\tau)=(1-\tau)^{-j-1}
\sum_{a=j-1}^{4j}
K_{a}^{(j)}\tau^{a}, \;
\forall j =1,{\ldots} , \infty \; .
\label{(101)}
\end{equation}
These formulae engender a $\zeta_{B}^{+}(0)$ which can be defined,
after change of variable from $z$ to $\tau$,
by splitting the integral with respect to $\tau$, in the integral
representation of $\zeta_{B}^{+}(s)$, according to the identity
$$
\int_{0}^{1}d\tau=\int_{0}^{\mu}d\tau+\int_{\mu}^{1}d\tau,
$$
and taking the limit as $\mu \rightarrow 1$ {\it after having evaluated 
the integral}. More precisely, since the integral on the left-hand side
is independent of $\mu$, we can choose $\mu$ small on the right-hand
side so that, in the interval $[0,\mu]$
(and only there!), we can use the uniform 
asymptotic expansion of the integrand where the negative powers of
$(1-\tau)$ are harmless. Moreover, independence of $\mu$ also
implies that, after having evaluated the integrals on the right-hand
side, we can take the $\mu \rightarrow 1$ limit. Within this framework,
the limit as $\mu \rightarrow 1$ of the second integral on the
right-hand side yields vanishing contribution to the asymptotic
expansion of $\zeta_{B}^{+}(s)$.

With this {\it caveat}, on defining (cf. (66)) 
\begin{equation}
Q_{\mu}(\alpha,\beta,\gamma) \equiv \int_{0}^{\mu}
\tau^{\alpha}(1-\tau)^{\beta}(1+\tau)^{\gamma} d\tau,
\label{(102)}
\end{equation}
we obtain the representations
\begin{eqnarray}
B_{+}(s)&=&-\omega_{0}(s){(\sin \pi s)\over \pi}\Bigr[
-Q_{\mu}(2s,-s-1,-s)+Q_{\mu}(2s,-s,-s-1) \nonumber \\
&-& Q_{\mu}(2s-1,-s,-s)\Bigr],
\label{(103)}
\end{eqnarray}
\begin{equation}
B_{j,+}(s)=-\omega_{j}(s){(\sin \pi s)\over \pi}
\sum_{a=j-1}^{4j}K_{a}^{(j)}Q_{\mu}(2s+a,-s-j-1,-s).
\label{(104)}
\end{equation}
The relevant properties of $Q_{\mu}(\alpha,\beta,\gamma)$ can be
obtained by observing that this function is nothing but a
hypergeometric function of two variables [\refcite{28}], i.e.
\begin{equation}
Q_{\mu}(\alpha,\beta,\gamma)={\mu^{\alpha+1}\over \alpha+1}
F_{1}(\alpha+1,-\beta,-\gamma,\alpha+2;\mu,-\mu).
\label{(105)}
\end{equation}
In detail, a summary of results needed to consider the limiting 
behaviour of $\zeta_{B}^{+}(s)$ as $s \rightarrow 0$ is
\begin{equation}
\omega_{0}(s){(\sin \pi s)\over \pi} \sim -5s+
{\rm O}(s^{2}),
\label{(106)}
\end{equation}
\begin{equation}
\omega_{j}(s){(\sin \pi s)\over \pi} \sim
{1\over 2}\delta_{j,3}+{\tilde b}_{j,1}s+{\rm O}(s^{2}),
\label{(107)}
\end{equation}
\begin{equation}
\lim_{\mu \to 1}Q_{\mu}(2s,-s-1,-s) \sim -{1\over s}+{\rm O}(s^{0}),
\label{(108)}
\end{equation}
\begin{equation}
\lim_{\mu \to 1}Q_{\mu}(2s,-s,-s-1) \sim \log(2)+{\rm O}(s),
\label{(109)}
\end{equation}
\begin{equation}
\lim_{\mu \to 1}Q_{\mu}(2s-1,-s,-s) \sim {1\over 2s}
+{\rm O}(s),
\label{(110)}
\end{equation}
\begin{eqnarray}
\; & \; & \lim_{\mu \to 1}Q_{\mu}(2s+a,-s-j-1,-s) \nonumber \\ 
&=&{\Gamma(-j-s)\Gamma(a+2s+1)\over \Gamma(a-j+s+1)}
{}_{2}F_{1}(a+2s+1,s,a-j+s+1;-1) \nonumber \\
&\sim &  {b_{j,-1}(a)\over s}+b_{j,0}(a)+{\rm O}(s),
\label{(111)}
\end{eqnarray}
where 
\begin{equation}
{\tilde b}_{j,1}=-1-2^{2-j}+\zeta_{R}(j-2)(1-\delta_{j,3})
+\gamma \delta_{j,3},
\label{(112)}
\end{equation}
\begin{equation}
b_{j,-1}(a)= {(-1)^{j+1}\over j!}
{\Gamma(a+1)\over \Gamma(a-j+1)}(1-\delta_{a,j-1}),
\label{(113)}
\end{equation}
and we only strictly need $b_{3,0}(a)$ which,
unlike the elliptic cases studied earlier,
now depends explicitly on $a$ and is given by ($\psi$ being the
standard notation for the logarithmic derivative of the 
$\Gamma$-function)
\begin{eqnarray}
b_{3,0}(a)&=& {1\over 6}{\Gamma(a+1)\over \Gamma(a-2)}\biggr[
-\log(2)-{1\over 4}(6a^{2}-9a+1){\Gamma(a-2)\over \Gamma(a+1)}
+ 2\psi(a+1) \nonumber \\
&-&\psi(a-2)-\psi(4)\biggr].
\label{(114)}
\end{eqnarray}

Remarkably, the coefficient of ${1\over s}$ in the small-$s$
behaviour of the generalized 
$\zeta$-function $\zeta_{B}^{+}(s)$ is zero because it is equal to
\begin{equation}
\lim_{s \to 0}s \zeta_{B}^{+}(s)=
\sum_{a=2}^{12}b_{3,-1}(a)K_{a}^{(3)}= {1\over 6}
\sum_{a=3}^{12}a(a-1)(a-2)K_{a}^{(3)},
\label{(115)}
\end{equation}
which vanishes by virtue of the rather peculiar general property
\begin{equation}
\sum_{a=j}^{4j}{\Gamma(a+1)\over \Gamma(a-j+1)}K_{a}^{(j)}
=\sum_{a=j}^{4j}\prod_{l=0}^{j-1}(a-l)K_{a}^{(j)}=0, \;
\forall j=1,{\ldots} ,\infty,
\label{(116)}
\end{equation}
and hence we find eventually
\begin{eqnarray}
\zeta_{B}^{+}(0)&=& -{5\over 4}+{1079\over 240}
+{5\over 2}-{1\over 2}\sum_{a=2}^{12}b_{3,0}(a)K_{a}^{(3)}
-\sum_{j=1}^{\infty}{\tilde b}_{j,1}
\sum_{a=j-1}^{4j}b_{j,-1}(a)K_{a}^{(j)} \nonumber \\
&=& {5\over 4}+{1079\over 240}+{599\over 720}
={296\over 45},
\label{(117)}
\end{eqnarray}
because the infinite sum on the first line of Eq. (117) vanishes 
by virtue of Eqs. (113) and (116), and exact cancellation of
$\log(2)$ terms is found to occur by virtue of Eq. (116).

To cross-check our analysis, we use Eq. (83) to evaluate
\begin{equation}
r_{k,+}(\tau)-r_{k,-}(\tau)={4\tau^{2}\over (1-\tau^{2})}u_{k-1}(\tau),
\label{(118)}
\end{equation}
and hence we find
\begin{equation}
R_{1,+}=R_{1,-}+{4\tau^{2}\over (1-\tau^{2})},
\label{(119)}
\end{equation}
\begin{equation}
R_{2,+}=R_{2,-}+{4\tau^{2}\over (1-\tau^{2})}
\left(u_{1}-{2\tau^{2}\over (1-\tau^{2})}-R_{1,-}\right),
\label{(120)}
\end{equation}
\begin{eqnarray}
R_{3,+}&=& R_{3,-}+{4\tau^{2}\over (1-\tau^{2})}
\left(u_{2}-{4\tau^{2}\over (1-\tau^{2})}u_{1}
-u_{1}R_{1,-}-R_{2,-}
+{4\tau^{2}\over (1-\tau^{2})}R_{1,-}\right) \nonumber \\
&+& {64\over 3}{\tau^{6}\over (1-\tau^{2})^{3}}
+{2\tau^{2}\over (1-\tau^{2})}R_{1,-}^{2},
\label{(121)}
\end{eqnarray}
and so on. This makes it possible to evaluate $B_{j,+}(s)-B_{j,-}(s)$
for all $j=1,2,... \infty$. Only $j=3$ contributes to $\zeta_{B}^{\pm}(0)$
(see below) and we find
\begin{eqnarray}
\; & \; & B_{3,+}(s)-B_{3,-}(s)=-\omega_{3}(s){(\sin \pi s)\over \pi}
\nonumber \\
& \cdot & \lim_{\mu \to 1} \int_{0}^{\mu}d\tau \; \tau^{2s}(1-\tau)^{-s}
(1+\tau)^{-s}{\partial \over \partial \tau}
(R_{3,+}-R_{3,-}).
\label{(122)}
\end{eqnarray}
The derivative in the integrand on the right-hand side of Eq. (122) reads 
as
\begin{equation}
{\partial \over \partial \tau}(R_{3,+}-R_{3,-})
=(1-\tau)^{-4}(1+\tau)^{-4}\Bigr(80\tau^{3}-24 \tau^{5}
+32 \tau^{7}-8 \tau^{9}\Bigr),
\label{(123)}
\end{equation}
and hence we can use again the definition (102) and the formula
(105) to express (122) through the functions 
$Q_{\mu}(2s+a,-s-4,-s-4)$, with $a=3,5,7,9$. This leads to
\begin{eqnarray}
\zeta_{B}^{+}(0)&=&\zeta_{B}^{-}(0)+B_{3,+}(0)-B_{3,-}(0) \nonumber \\
&=& \zeta_{B}^{-}(0)-{1\over 24}\sum_{l=1}^{4}
{\Gamma(l+1)\over \Gamma(l-2)}\left[\psi(l+2)-{1\over (l+1)}\right]
\kappa_{2l+1}^{(3)} \nonumber \\
&=&{206\over 45}+2={296\over 45},
\label{(124)}
\end{eqnarray}
where $\kappa_{2l+1}^{(3)}$ are the four coefficients on the
right-hand side of (123).
Regularity of $\zeta_{B}^{+}(s)$ at the origin is guaranteed because
$\lim_{s \to 0}s \zeta_{B}^{+}(s)$ is proportional to
$$
\sum_{l=1}^{4}{\Gamma(l+1)\over \Gamma(l-2)}\kappa_{2l+1}^{(3)}=0,
$$
which is a particular case of the peculiar spectral cancellation (cf. (116))
\begin{equation}
\sum_{a=a_{\rm min}(j)}^{a_{\rm max}(j)}
{\Gamma \left({(a+1)\over 2}\right)\over 
\Gamma \left({(a+1)\over 2}-j \right)}
\kappa_{a}^{(j)}=0,
\label{(125)}
\end{equation}
where $a$ takes both odd and even values. The case $j=3$ is simpler because
then only $\kappa_{a}^{(j)}$ coefficients with odd $a$ are non-vanishing.

Remaining contributions to $\zeta(0)$,
being obtained from strongly elliptic sectors of the
boundary-value problem, are easily found to agree with the
results in Ref. 19, i.e.
\begin{equation}
\zeta(0)[{\rm transverse} \; 
{\rm traceless} \; {\rm modes}]=-{278\over 45},
\label{(126)}
\end{equation}
\begin{equation}
\zeta(0)[{\rm coupled} \; {\rm vector} 
\; {\rm modes}]={494\over 45},
\label{(127)}
\end{equation}
\begin{equation}
\zeta(0)[{\rm decoupled} \; {\rm vector} 
\; {\rm mode}]=-{15\over 2},
\label{(128)}
\end{equation}
\begin{equation}
\zeta(0)[{\rm scalar} \; {\rm modes}(a_{1},e_{1};a_{2},b_{2},e_{2})]
=-17,
\label{(129)}
\end{equation}
\begin{equation}
\zeta(0)[{\rm scalar} \; {\rm ghost} 
\; {\rm modes}]=-{149\over 45},
\label{(130)}
\end{equation}
\begin{equation}
\zeta(0)[{\rm vector} \; {\rm ghost} 
\; {\rm modes}]={77\over 90},
\label{(131)}
\end{equation}
\begin{equation}
\zeta(0)[{\rm decoupled} \; {\rm ghost} \; {\rm mode}]={5\over 2}.
\label{(132)}
\end{equation}
Our full $\zeta(0)$ is therefore, from (79), (80), (94), (117),
(126)-(132), $\zeta(0)={142\over 45}$.

\section{Concluding remarks}

We have studied the
analytically continued eigenvalue conditions for metric perturbations
on the Euclidean four-ball, in the presence of boundary conditions 
completely invariant under infinitesimal diffeomorphisms in the de
Donder gauge and with the $\alpha$ parameter set to $1$ in Eq. (10). 
This has made it possible to prove
that only one sector of the scalar-mode
determinant is responsible for lack of strong ellipticity of the
boundary-value problem (see second line of Eq. (44) and the analysis
in Secs. 3 and 4). The first novelty with respect to the work in
Ref. 18 is a clear separation of the elliptic and
non-elliptic sectors of spectral asymptotics for Euclidean 
quantum gravity. We have also shown that one can indeed obtain a
regular $\zeta$-function asymptotics at small $s$ in
the non-elliptic case by virtue of the
remarkable identity (116). Our prescription for the
$\zeta(0)$ value differs from  
the result first obtained in
Ref. 19, where, however, neither the strong ellipticity
issue [\refcite{18}] nor the non-standard spectral asymptotics of
our Sec. 4 had been considered.

As far as we can see, the issues raised by our results are as follows.
\vskip 0.3cm
\noindent
(i) The integral representation (81) is legitimate because the
second line of Eq. (44) corresponds to the eigenvalue conditions,
for $n \geq 3$,
\begin{equation}
F_{B}^{\pm}(n,x) \equiv J_{n}'(x)+\left(-{x\over 2} \pm 
{n\over x} \right)J_{n}(x)=0.
\label{(133)}
\end{equation}
For both choices of sign in front of ${n\over x}$, if $x_{i}$ is a root,
then so is $-x_{i}$, with positive eigenvalue $E_{i}=x_{i}^{2}$
(having set the 3-sphere radius $q=1$ for simplicity). For any fixed
$n$, there is a countable infinity of roots $x_{i}$ and 
they grow approximately linearly with the integer $i$ counting such roots. 
The function $F_{B}^{\pm}$ admits therefore a canonical-product representation
(Ahlfors [\refcite{29}]) which ensures that the integral representation (81)
reproduces the standard definition of generalized $\zeta$-function, i.e.
$$
\zeta(s) \equiv \sum_{E_{k}>0}d(E_{k})E_{k}^{-s},
$$
where $d(E_{k})$ is the degeneracy of the eigenvalue $E_{k}$.
\vskip 0.3cm
\noindent
(ii) Even though the lack of strong ellipticity implies that the
functional trace of the heat semigroup no longer exists, and hence
the Mellin transform relating $\zeta$-function to integrated 
heat kernel cannot be exploited, it remains possible to define the
functional determinant of the operator $P$ acting on metric
perturbations. For this purpose, a weaker assumption provides a 
sufficient condition, i.e. the existence of a sector in the complex
plane free of eigenvalues of the leading symbol of $P$ 
(Seeley [\refcite{30}]).
Note also that, if one looks at the $A_{1}$ heat-kernel coefficient for
boundary conditions involving tangential 
derivatives [\refcite{8}], it is
exactly for the ball that the potentially divergent pieces involving 
the extrinsic curvature in $A_{1}$ cancel. Thus, on the Euclidean ball
cancellations take place that maybe could explain why $\zeta(0)$ is
finite. This might be therefore {\it a very particular result for the ball}.
\vskip 0.3cm
\noindent
(iii) By virtue of standard recurrence relations among Bessel functions, 
the eigenvalue conditions (133) are equivalent to studying the eigenvalue
conditions 
\begin{equation}
{\widetilde F}_{B}^{\pm}(n,x)=J_{n}(x) \mp {2\over x}J_{n-1}(x)=0,
\label{(134)}
\end{equation}
where the eigenvalues $E(i,n,\pm)$ are obtained by squaring up the roots
$x(i,n,\pm)$. The equation for ${\widetilde F}_{B}^{-}(n,x)$ can be
further re-expressed in the form
\begin{equation}
\left(1+{4n \over x^{2}}\right)J_{n}(x)-{2\over x}J_{n-1}(x)=0.
\label{(135)}
\end{equation}
The functions ${\widetilde F}_{B}^{\pm}$ differ therefore by one term
only, and this term gets small as $x$ gets larger. The numerical analysis
confirms indeed that a $\rho(i,n)$ positive and much smaller than $1$
exists such that one can write (Esposito et al. [\refcite{31}])
\begin{equation}
E(i,n,+)=E(1,n,+)\delta_{i,1}+E(i-1,n,-)(1+\rho(i,n))(1-\delta_{i,1}),
\label{(136)}
\end{equation}
for all $n \geq 3$ and for all $i \geq 1$. 
\vskip 0.3cm
\noindent
(iv) The remarkable factorization of eigenvalue conditions, with
resulting isolation of elliptic part of spectral asymptotics
(transverse-traceless, vector and ghost modes, all modes in
finite-dimensional sub-spaces and three of the four equations for
scalar modes), suggests trying to re-assess functional integrals on
manifolds with boundary, with the hope of being able to obtain
unique results from the
non-elliptic contribution. If this cannot be achieved, 
the two alternatives below 
should be considered again.
\vskip 0.3cm
\noindent
(v) Luckock boundary conditions (Luckock [\refcite{32}]),
which engender BRST-invariant amplitudes but are not 
diffeomorphism invariant [\refcite{15}]. They have already been applied
by Moss and Poletti [\refcite{12,33}]. 
\vskip 0.3cm
\noindent
(vi) Non-local boundary conditions that lead to
surface states in quantum cosmology and pseudo-differential 
operators on metric and ghost modes (Marachevsky and Vassilevich, 
Esposito [\refcite{34}]). Surface states are
particularly interesting since they describe a transition from quantum
to classical regime in cosmology entirely ruled by the strong ellipticity
requirement, while pseudo-differential operators 
are a source of technical complications.

There is therefore encouraging evidence in favour of Euclidean 
quantum gravity being able to drive further developments in
quantum field theory, quantum cosmology and spectral asymptotics
(see early mathematical papers by Grubb [\refcite{35}], Gilkey and
Smith [\refcite{36}]) in the years to come.

\section*{Appendix: Olver expansions}

In Secs. 3 and 4 we use the uniform asymptotic expansion of modified
Bessel functions $I_{\nu}$ first found by Olver [\refcite{37}]:
\begin{equation}
I_{\nu}(z\nu) \sim {e^{\nu \eta}\over \sqrt{2\pi \nu}
(1+z^{2})^{1\over 4}}\left(1+\sum_{k=1}^{\infty}
{u_{k}(\tau)\over \nu^{k}}\right),
\label{(137)}
\end{equation}
where
\begin{equation}
\tau \equiv (1+z^{2})^{-{1\over 2}}, \;
\eta \equiv (1+z^{2})^{1\over 2}
+{\rm log}\left({z \over 1+\sqrt{1+z^{2}}}\right).
\label{(138)}
\end{equation}
This holds for $\nu \rightarrow \infty$ at fixed $z$. The polynomials
$u_{k}(\tau)$ can be found from the recurrence relation [\refcite{27}]
\begin{equation}
u_{k+1}(\tau)={1\over 2}\tau^{2}(1-\tau^{2})u_{k}'(\tau)
+{1\over 8}\int_{0}^{\tau}d\rho \; (1-5\rho^{2})u_{k}(\rho),
\label{(139)}
\end{equation}
starting with $u_{0}(\tau)=1$. Moreover, the first derivative of
$I_{\nu}$ has the following uniform asymptotic expansion at large $\nu$ 
and fixed $z$:
\begin{equation}
I_{\nu}'(z \nu) \sim {e^{\nu \eta}\over \sqrt{2 \pi \nu}}
{(1+z^{2})^{1\over 4}\over z}\left(1+\sum_{k=1}^{\infty}
{v_{k}(\tau)\over \nu^{k}}\right),
\label{(140)}
\end{equation}
with the $v_{k}$ polynomials determined from the $u_{k}$ 
according to [\refcite{27}]
\begin{equation}
v_{k}(\tau)=u_{k}(\tau)+\tau(\tau^{2}-1)\left[
{1\over 2}u_{k-1}(\tau)+\tau u_{k-1}'(\tau)\right],
\label{(141)}
\end{equation}
starting with $v_{0}(\tau)=u_{0}(\tau)=1$.

\section*{Acknowledgments}
We are grateful to Bernhelm Boo\ss--Bavnbek and Krysztof
Wojciechowski for their kind invitation and encouragement. Moreover,
we are indebted to Ivan Avramidi for inspiration provided by
previous collaboration with some of us and by continuous
correspondence, and to Gerd Grubb for enlightening correspondence.
K. Kirsten is grateful to the Baylor University Research 
Committee, to the Max-Planck-Institute for Mathematics in the
Sciences (Leipzig, Germany) and to the 
INFN for financial support. The 
work of G. Esposito and K. Kirsten has been partially supported
also by PRIN {\it SINTESI}. The work of A.Yu. Kamenshchik 
was partially supported
by the Russian Foundation for Basic Research under the Grant No. 
02-02-16817 and by the Scientific School Grant No. 2338.2003.2

\end{document}